\RequirePackage{lineno}
\documentclass[twocolumn,aps,showpacs,superscriptaddress,floatfix]{revtex4}
\usepackage{epsfig}
\usepackage{color}
\newcommand {\snn}	{\sqrt{s_{_{\rm NN}}}}
\newcommand {\gevc}	{GeV/$c$}
\newcommand {\fmc}	{fm/$c$}
\newcommand {\dAu}	{$d$+Au}
\newcommand {\pbar}	{\bar{p}}
\newcommand {\pt}	{p_{\perp}}
\newcommand {\rt}	{r_{\perp}}
\newcommand {\psir}	{\psi_2^{(r)}}
\newcommand {\phirndm}	{$\phi$-randomized}
\newcommand {\mean}[1]	{\langle #1\rangle}

\begin{document}
\title{Origin of the mass splitting of elliptic anisotropy in a multiphase transport model}
\author{Hanlin Li}
\affiliation{College of Science, Wuhan University of Science and Technology, Wuhan, Hubei 430065, China}
\affiliation{Department of Physics and Astronomy, Purdue University, West Lafayette, Indiana 47907, USA}
\author{Liang He}
\affiliation{Department of Physics and Astronomy, Purdue University, West Lafayette, Indiana 47907, USA}
\author{Zi-Wei Lin}
\affiliation{Department of Physics, East Carolina University, Greenville, North Carolina 27858, USA}
\author{Denes Molnar}
\affiliation{Department of Physics and Astronomy, Purdue University, West Lafayette, Indiana 47907, USA}
\author{Fuqiang Wang}
\affiliation{Department of Physics and Astronomy, Purdue University, West Lafayette, Indiana 47907, USA}
\author{Wei Xie}
\affiliation{Department of Physics and Astronomy, Purdue University, West Lafayette, Indiana 47907, USA}
\date{\today}

\begin{abstract}
The mass splitting of elliptic anisotropy ($v_2$) at low transverse momentum is considered as a hallmark of hydrodynamic collective flow. We investigate a multiphase transport (AMPT) model where the $v_2$ is mainly generated by an anisotropic escape mechanism, not of the hydrodynamic flow nature, and where mass splitting is also observed. We demonstrate that the $v_2$ mass splitting in AMPT is small right  after hadronization (especially when resonance decays are included); the mass splitting mainly comes from hadronic rescatterings, even though their contribution to the overall charged hadron $v_2$ is small. These findings are qualitatively the same as those from hybrid models that combine hydrodynamics with a hadron cascade. We further show that there is no qualitative difference between heavy ion collisions and small  system collisions. Our results indicate that the $v_2$ mass splitting is not a unique signature of hydrodynamic collective flow and thus cannot distinguish whether the elliptic flow is generated mainly from hydrodynamics or the anisotropic parton escape.
\end{abstract}
\maketitle

{\em Introduction.}
Relativistic heavy ion collisions aim to create quark-gluon plasma (QGP) to allow study of quantum chromodynamics (QCD) at the extreme conditions of high temperature and energy density~\cite{Arsene:2004fa,Back:2004je,Adams:2005dq,Adcox:2004mh,Muller:2012zq}. The system created in these collisions is described well by hydrodynamics where the high pressure buildup drives the system to expand at relativistic speed~\cite{Heinz:2013th,Gale:2013da}. Experimental data fit with hydrodynamics inspired models are consistent with particles being locally thermalized with a common radial flow velocity~\cite{Abelev:2008ab}. Of particular interest are non-central collisions where the overlap zone of the colliding nuclei is anisotropic in the transverse plane (perpendicular to beam). The pressure gradient would generate anisotropic expansion and final-state elliptic flow~\cite{Ollitrault:1992bk}. Large elliptic anisotropy in momentum ($v_2$) has been measured, as large as hydrodynamic calculations predict~\cite{Arsene:2004fa,Heinz:2013th,Gale:2013da}. This suggests that the collision system is a strongly interacting and nearly thermalized QGP, dubbed sQGP~\cite{Gyulassy:2004zy}. 

A hallmark of the hydrodynamic description of relativistic heavy ion collisions is the mass splitting of $v_2$ at a given low transverse momentum ($\pt$)~\cite{Huovinen:2001cy,Heinz:2013th}. It is consistent with a common radial velocity field, whose azimuthal modulation gives rise to momentum-space azimuthal anisotropy, and whose effect on hadron $\pt$ via the Cooper-Frye hadronization mechanism~\cite{Cooper:1974mv} (commonly exploited in hydrodynamic calculations) gives rise to the mass splitting. Results from hybrid models, where hydrodynamics is followed by a hadron cascade, have shown that the $v_2$ mass splitting is small just after hadronization when resonance decays are included and that the mass splitting is strongly enhanced by hadronic scatterings~\cite{Hirano:2007ei,Song:2010aq,Romatschke:2015dha}. It has also been shown that the magnitude of the mass splitting from the hydrodynamical stage alone depends strongly on the kinetic freeze-out temperature~\cite{Hirano:2007ei}. 

It is generally perceived that large $v_2$ can only be generated in large-system heavy ion collisions, and hydrodynamics is a highly plausible scenario for how the collision system evolves. Recent particle correlation data, however, hint at similar $v_2$ and mass splitting effects in small systems of high multiplicity 
$p$+$p$ and $p$+Pb collisions at the Large Hadron Collider~\cite{Khachatryan:2010gv,CMS:2012qk,Abelev:2012ola,Aad:2012gla} and \dAu\ collisions at the Relativistic Heavy Ion Collider (RHIC)~\cite{Adare:2014keg,Adamczyk:2015xjc}. Hydrodynamics has been applied to these systems and seems to successfully describe the experimental data, including the mass splitting~\cite{Bozek:2010pb,Bozek:2012gr}. This could suggest that these small-system collisions create a sQGP as well, in contrast to general expectation. 

On the other hand, parton transport models, such as a multiphase transport model (AMPT)~\cite{Zhang:1999bd,Lin:2001zk,Lin:2004en}, have also been widely used to describe experimental data. The string melting version of AMPT~\cite{Lin:2001zk,Lin:2004en} reasonably reproduces particle yields, $\pt$ spectra, and $v_2$ of low-$\pt$ pions and kaons in central and mid-central Au+Au collisions at $200A$~GeV and Pb+Pb collisions at $2760A$~GeV~\cite{Lin:2014tya}. The small system data can be also satisfactorily described by AMPT~\cite{Bzdak:2014dia}. The successful description by AMPT of experimental data, especially the heavy ion data, did not come as a surprise, because it has been thought that the transport models have approached hydrodynamic limit due to high energy densities and/or large parton-parton interaction cross sections. 

However, a recent study using AMPT shows that the azimuthal anisotropy is mainly generated by anisotropic parton escape from the collision zone~\cite{He:2015hfa,Lin:2015ucn}; hydrodynamics may play only a minor role. The escape mechanism would naturally explain the measured azimuthal anisotropies in both heavy ion and small system collisions. While the escape mechanism does not generate radial flow, $v_2$ mass splitting is also present in AMPT. This suggests that hydrodynamic radial flow may not be the only mechanism that can generate the mass splitting of $v_2$~\cite{Borghini:2010hy}. The underlying reason for the mass splitting in AMPT is the question we address in this study.

{\em Model and Analysis.}
We employ the same string melting version of AMPT~\cite{Lin:2001zk,Lin:2004en} (v2.26t5, available online~\cite{ampt}) as in our earlier study~\cite{He:2015hfa}. The model consists of fluctuating initial conditions, $2\to2$ parton elastic scatterings~\cite{Zhang:1997ej}, quark coalescence for hadronization, and hadronic interactions. We use Debye screened differential cross-section $d\sigma/dt\propto\alpha_s^2/(t-\mu_D^2)^2$~\cite{Lin:2004en}, with strong coupling constant $\alpha_s=0.33$ and Debye screening mass $\mu_D=2.265$/fm. The total parton scattering cross section is then $\sigma=3$~mb for all AMPT calculations in this study. After partons stop interacting, a simple quark coalescence model is applied to combine two nearest partons into a meson and three nearest quarks (or antiquarks) into a baryon (or an antibaryon)~\cite{Lin:2004en}. Subsequent interactions of these formed hadrons are modeled by a hadron cascade, which explicitly includes particles such as $\pi$, $\rho$, $\omega$, $\eta$, $K$, $K^*$, $\phi$ mesons, $N$, $\Delta$, $N^*(1440)$, $N^*(1535)$, $\Lambda$, $\Sigma$, $\Xi$, $\Omega$ baryons and antibaryons~\cite{Lin:2004en}, plus deuterons and anti-deuterons~\cite{Oh:2009gx}. We terminate the hadronic interactions at a cutoff time, when the observables of interest are stable; a cutoff time of 30~\fmc\ is used.

We simulate Au+Au collisions (impact parameter $b=6.6$-8.1~fm corresponding to 20-30\% centrality~\cite{Abelev:2008ab}) and \dAu\ collisions at $b=0$~fm at $200A$~GeV using AMPT. The AMPT version and parameter values are the same as those used for RHIC collisions in earlier studies~\cite{Lin:2014tya,He:2015hfa}. We compute the 2nd harmonic plane of each event from its initial configuration of all partons~\cite{Ollitrault:1993ba} via
\begin{equation}
\psir=\left[{\rm atan2}(\mean{\rt^{2}\sin2\phi_r},\mean{\rt^{2}\cos2\phi_r})+\pi\right]/2\,.
\end{equation}
Here $\rt$ and $\phi_r$ are the polar coordinate of each initial parton (after its formation time) in the transverse plane, and $\mean{...}$ denotes the per-event average. We analyze the momentum-space azimuthal anisotropy of partons in the final state before hadronization, of hadrons right after hadronization before hadronic rescatterings take place, and of freeze-out hadrons in the final state. The momentum anisotropy is characterized by Fourier coefficients~\cite{Voloshin:1994mz}
\begin{equation}
v_2=\mean{\cos n(\phi -\psir)}\,,
\end{equation}
where $\phi$ is the azimuthal angle of the particle (parton or hadron) momentum. The resolution (accuracy) of $\psir$ is practically 100\% due to the large initial parton multiplicity~\cite{Xiao:2012uw}. All results shown in this study are for particles within pseudo-rapidity window $|\eta|<1$.

\begin{figure}[hbt]
  \begin{center}
    \includegraphics[width=0.9\columnwidth]{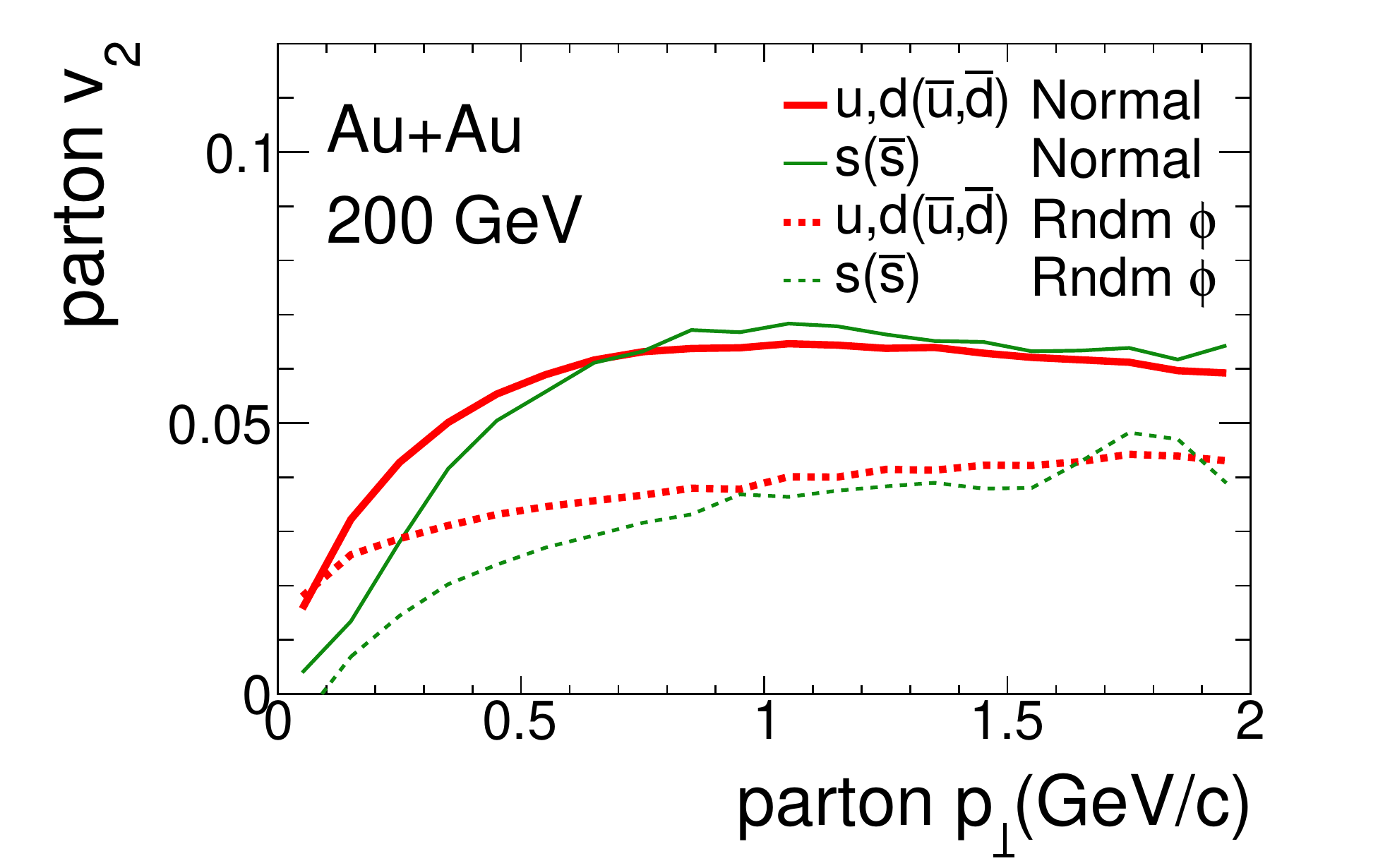}
    \caption{{\em Parton $v_2$.} Mid-rapidity ($|\eta|<1$) parton $v_2$ as a function of $\pt$ for $u$+$d$ (thick curves) and $s$ (anti)quarks (thin curves) in the final state before hadronization in $b=6.6$-8.1~fm Au+Au collisions at $\snn=200$~GeV by normal AMPT (solid) and \phirndm\ AMPT (dashed). }
    \label{fig:uds}
  \end{center}
\end{figure}

{\em Partonic anisotropy.}
AMPT has only quarks but no gluons. The gluon degrees of freedom can be considered as being absorbed in the quarks. Figure~\ref{fig:uds} shows the elliptic anisotropies of the $u$ and $d$ light (anti)quarks and the $s$ strange (anti)quarks by the solid curves. 
We find practically identical $v_2$ values for quarks and antiquarks of the same flavor, and for $u(\bar{u}$) and $d(\bar{d}$) quarks as well. Therefore, we only plot light quark ($u,d,\bar{u},\bar{d}$ combined) and strange quark ($s,\bar{s}$) $v_2$.
At low $\pt$ the light quark $v_2$ is larger than the $s$ quark $v_2$. 
Although $v_2$ comes largely from the anisotropic escape mechanism, there does exist a contribution from hydrodynamics in AMPT~\cite{He:2015hfa}. We thus also carry out a test calculation with no collective anisotropic flow by randomizing the outgoing parton azimuthal directions after each parton-parton scattering as in Ref.~\cite{He:2015hfa}. Since the parton azimuthal angles are randomized, the final-state parton anisotropy is entirely due to the anisotropic escape mechanism~\cite{He:2015hfa}. However, difference between light and $s$ quark $v_2$'s is still observed as shown by the dashed curves. The fact that this difference is similar between the normal and \phirndm\ AMPT~\cite{prc} suggests that it may be caused by kinematic difference in the scattering processes due to their mass difference rather than collective flow. At high $\pt$ their $v_2$'s approach to each other as expected because the mass difference becomes unimportant (and quark scattering cross-sections are all set to be the same).

\begin{figure}[hbt]
  \begin{center}
    \includegraphics[width=0.9\columnwidth]{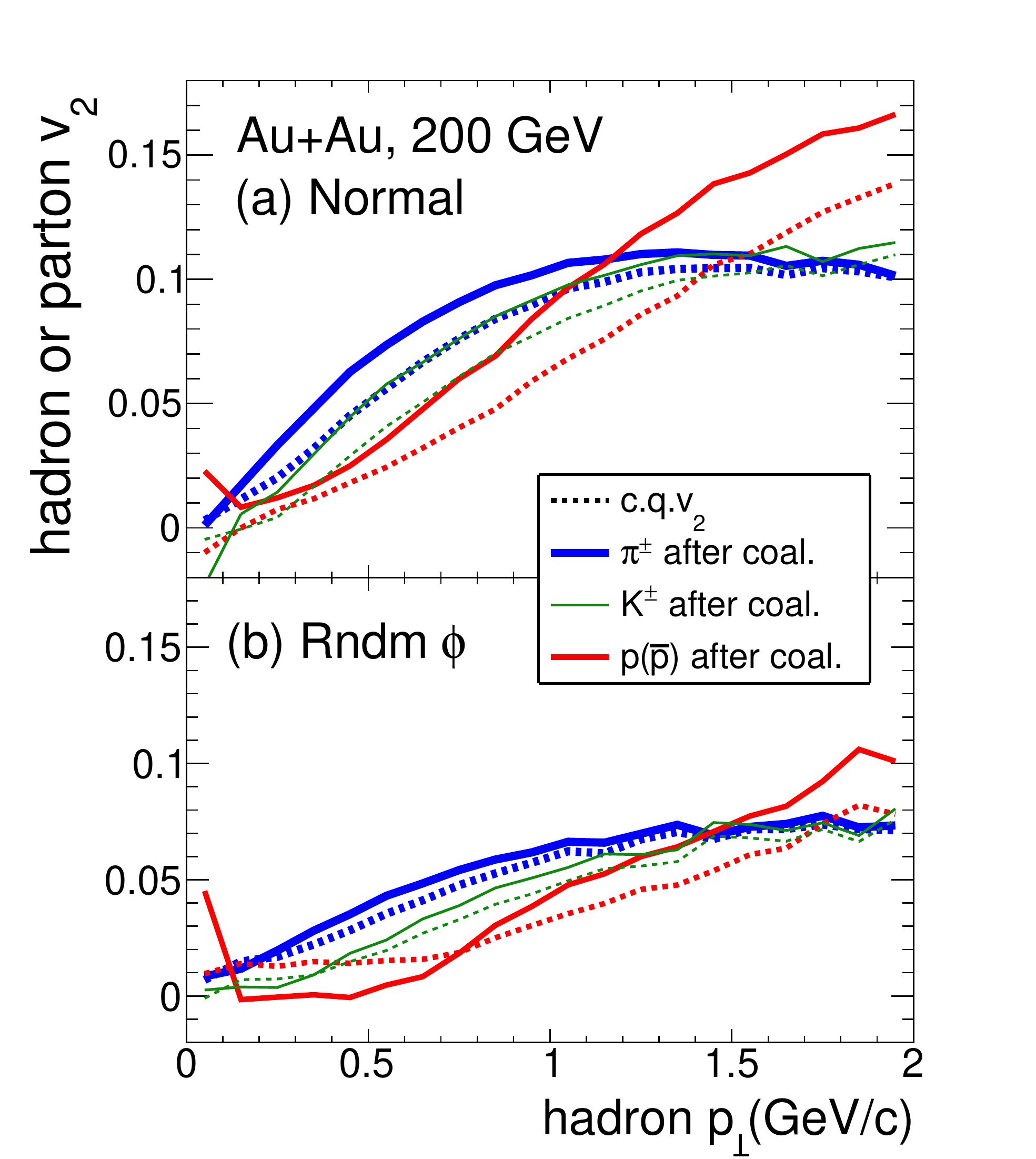}
    \caption{{\em Mass splitting from coalescence.} Mid-rapidity ($|\eta|<1$) constituent quark (c.q., dashed curves) and {\em primordial} hadron (solid curves) $v_2$ both as a function of {\em hadron} $\pt$ right after hadronization before hadronic rescatterings in $b=6.6$-8.1~fm Au+Au collisions at $\snn=200$~GeV by normal AMPT (a) and \phirndm\ AMPT (b).}
    \label{fig:coal}
  \end{center}
\end{figure}

{\em Mass splitting from coalescence.}
Since primordial (i.e.~directly formed by quark coalescence) pions and protons all come from light quarks, the difference between their $v_2$'s must come from the hadronization process and/or hadronic rescatterings. We study the effect of the former by examining $v_2$ of hadrons right after hadronization, before any hadronic rescatterings take place. Figure~\ref{fig:coal}(a) shows the $v_2$ of {\em primordial} $\pi^{\pm}$, $K^{\pm}$, and (anti)proton as a function of $\pt$ (solid curves) in Au+Au collisions at $\snn=200$~GeV. The difference observed between pion and proton $v_2$ can only come from the hadronization process. The string melting AMPT model forms hadrons via quark coalescence~\cite{Lin:2009tk}. The pion and proton $v_2$ difference arises from the different numbers of constituent quarks they are made of and the kinematics of those coalescing (anti)quarks. At high $\pt$ the hadron $v_2$ has been measured to exhibit a scaling in the number of constituent quarks after the hadron $\pt$ is divided by the number of constituent quarks, $v_2^{\rm baryon}(\pt/3)/3 \approx v_2^{\rm meson}(\pt/2)/2$. This has a natural explanation~\cite{Molnar:2003ff} in quark coalescence, where two or three relatively high $\pt$ quarks are more or less collimated and coalesce into a meson or baryon. Mesons and baryons, respectively, take on two and three times the quark $v_2$ (which are saturated at high $\pt$ as seen in Fig.~\ref{fig:uds}). Apparently, this quark coalescence picture cannot be extended to low $\pt$~\cite{Lin:2009tk}; if it could be, then, because the quark $v_2$ is approximately linear at low $\pt$ (see Fig.~\ref{fig:uds}), the meson and baryon $v_2(\pt)$ as a function of $\pt$ would coincide with each other (the two or three constituent quark $\pt$'s add to the hadron $\pt$ and the quark $v_2$'s also add to the hadron $v_2$) and there would be no mass splitting. The mass splitting of $v_2$ at low $\pt$ in AMPT comes from the dynamics in the coalescence process~\cite{Lin:2009tk}, such as the finite opening angles or kinematics~\cite{prc}. The dynamical ``selections'' of constituent quarks into pions, kaons, and protons are manifest in the constituent quark $v_2$ distributions shown by the dashed curves in Fig.~\ref{fig:coal}(a), plotted at the respective {\em hadron} $\pt$. The constituent quarks for a given hadron $\pt$ value are spread in their parton $\pt$'s and their $v_2$'s do not arithmetically add up to the hadron $v_2$ because of finite opening angles~\cite{prc}.

Figure~\ref{fig:coal}(b) shows the $v_2$ results from \phirndm\ AMPT~\cite{He:2015hfa} for {\em primordial} hadrons right after coalescence hadronization and the corresponding constituent quark $v_2$'s. No hydrodynamic anisotropic flow is present in the \phirndm\ case~\cite{He:2015hfa}, however, mass splitting is still observed. This reinforces our conclusion that the mass splitting is mainly due to kinematics in the coalescence process; it is therefore not a unique signature of collective anisotropic flow or hydrodynamics.

\begin{figure}[hbt]
  \begin{center}
    \includegraphics[width=0.9\columnwidth]{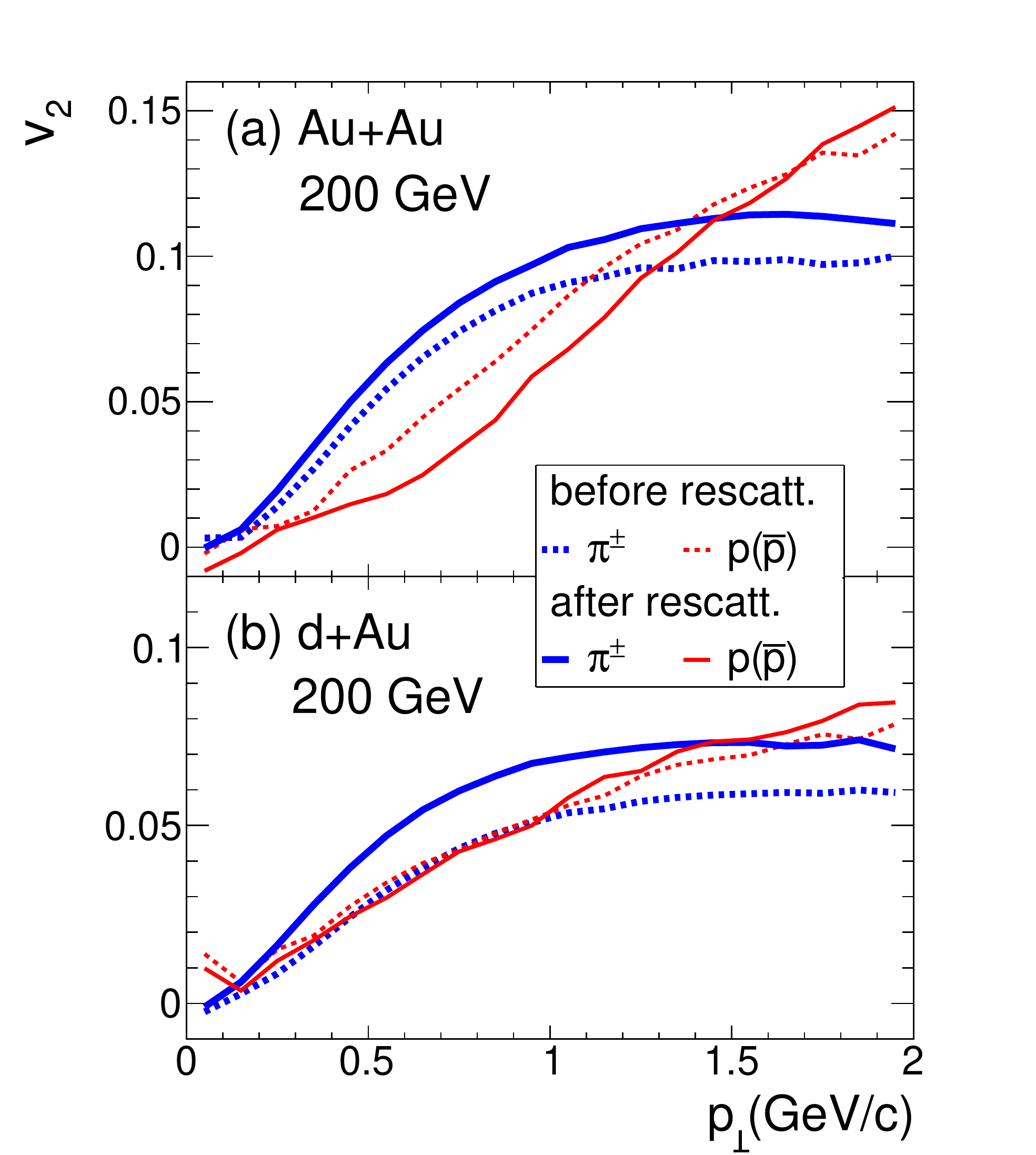}
    \caption{{\em Effects of hadronic rescatterings.} Mid-rapidity ($|\eta|<1$) pion and (anti)proton $v_2$ before (dashed) and after (solid) hadronic rescatterings in $b=6.6$-8.1~fm Au+Au (a) and $b=0$~fm \dAu\ (b) collisions at $\snn=200$~GeV by AMPT. Resonance decays are included.}
    \label{fig:rescatt}
  \end{center}
\end{figure}

{\em Effects of hadronic rescatterings.}
After hadronization, particles interact both inelastically and elastically. Unstable particles decay. Measured in detectors are particles after interactions cease and after resonances decay. In order to study effects of hadronic rescatterings on $v_2$, we ought to first evaluate the $v_2$ of hadrons before hadronic rescatterings but including effects of resonance decays. To this end, we run AMPT with the maximum time for hadronic scatterings set to 0.6~\fmc, and then analyze $v_2$ of the ``final-state'' hadrons. The final-state hadrons from such a setting do not have hadronic rescatterings but include decay products. The results are shown by the dashed curves in Fig.~\ref{fig:rescatt}(a) for Au+Au collisions. The decay product $v_2$ is usually smaller than their parent $v_2$~\cite{prc}. By including decay products, the hadron $v_2$'s (dashed curves in Fig.~\ref{fig:rescatt}(a)) are reduced from those of primordial hadrons (solid curves in Fig.~\ref{fig:coal}(a)); so is the magnitude of mass-splitting in $v_2$.

The $v_2$ values before hadronic rescatterings (including resonance decay effects) can be considered as the initial $v_2$ for the hadronic evolution stage. The final-stage freezeout hadron $v_2$'s (also including decay daughters) are shown in Fig.~\ref{fig:rescatt}(a) by the solid curves. Pion $v_2$ increases, while proton $v_2$ decreases from before to after hadronic rescatterings. This may be understood as follows. Because of interactions between pions and protons, they tend to flow together at the same velocity. Thus, the same-velocity pions and protons (i.e.~small $\pt$ pions and large $\pt$ protons) will tend to have the same anisotropy. This will yield lower $v_2$ for the protons and higher $v_2$ for the pions at the same $\pt$ value. This should happen whether or not there is a net gain in the overall charged hadron $v_2$, which depends on the initial configuration geometry from which the hadronic evolution begins~\cite{prc}.

Figure~\ref{fig:rescatt}(b) shows the results for \dAu\ collisions. There, pion $v_2$ increases significantly due to hadronic rescatterings, while the proton $v_2$ remains roughly unchanged. This is a net effect of the splitting (i.e.~increase in pion $v_2$ and decrease in proton $v_2$) due to pion-proton interactions and an overall gain of $v_2$ for charged hadrons. The additional gain in the charged hadron $v_2$ is larger in \dAu\ than Au+Au collisions, and this is due to a larger eccentricity in the \dAu\ system at the start of hadron cascade~\cite{prc}. 

\begin{figure}[hbt]
  \begin{center}
    \includegraphics[width=0.9\columnwidth]{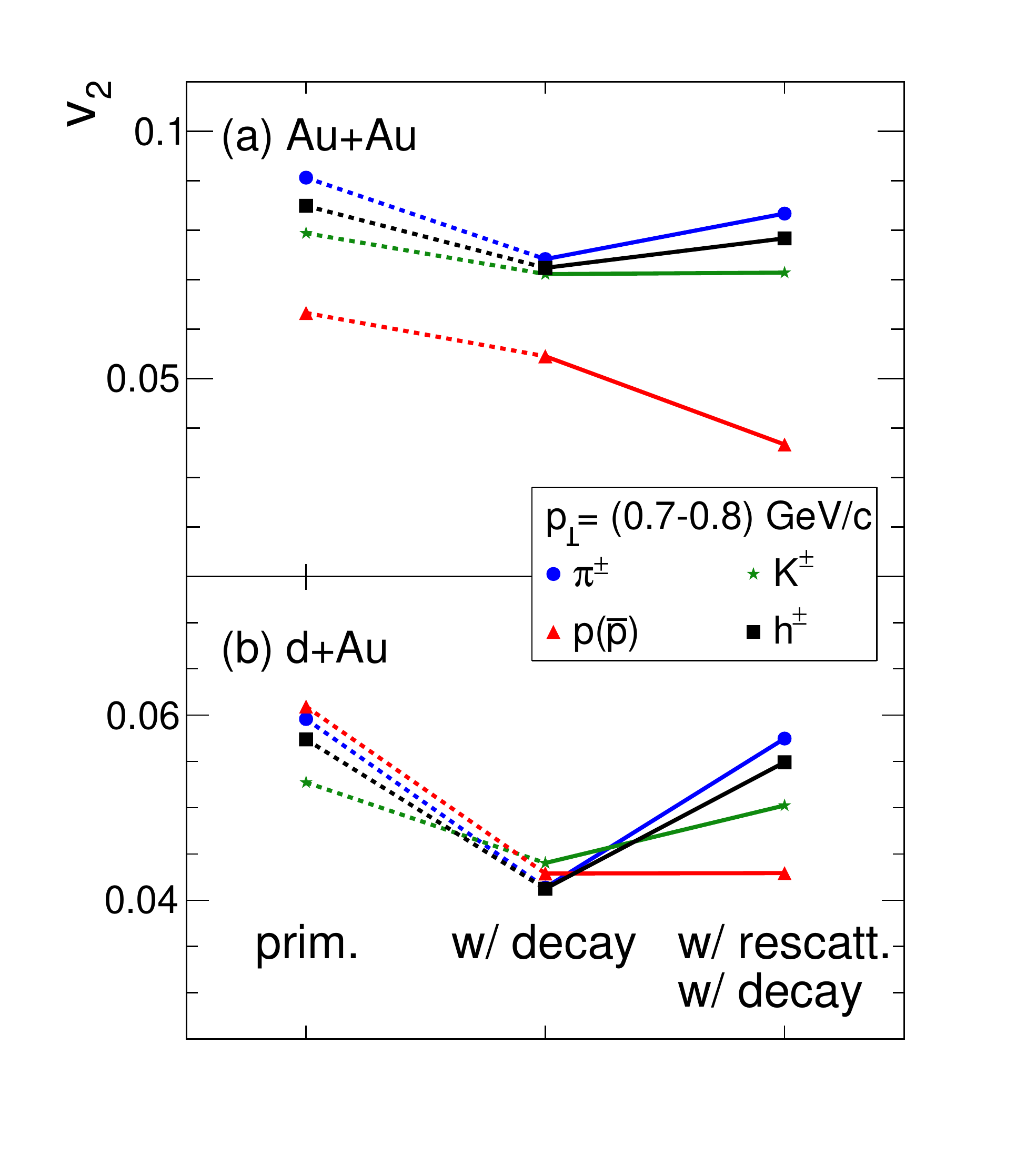}
    \vspace*{-5mm}
    \caption{{\em Origins of $v_2$ mass splitting.} Mid-rapidity ($|\eta|<1$) $v_2$ of $\pi^{\pm}$, $K^{\pm}$, $p(\pbar$) and charged hadrons ($h^\pm$) at $0.7<\pt<0.8$~\gevc\ at different stages of system evolution in $b=6.6$-8.1~fm Au+Au (a) and $b=0$~fm \dAu\ (b) collisions at $\snn=200$~GeV by AMPT: right after coalescence hadronization and before hadronic rescatterings (initial $v_2$ of primordial hadrons), hadron initial $v_2$ including decays, and after hadronic rescatterings at freezeout (hadron final $v_2$).}
    \label{fig:summary}
  \end{center}
\end{figure}

To summarize the origin of $v_2$ mass splitting, we plot in Fig.~\ref{fig:summary} the $v_2$ of pions, kaons, and protons within $0.7<\pt<0.8$~\gevc, a typical region where the mass splitting is manifest, for different stages of the collision system evolution: (i) right after coalescence hadronization including only primordial hadrons (labeled ``prim.''); (ii) right after coalescence hadronization but including decay products (labeled ``w/ decay''); and (iii) at final freezeout (labeled ``w/ rescatt.~w/ decay). As shown in Fig.~\ref{fig:summary}, most of the final-state hadron $v_2$ is built up in the partonic phase (more so in Au+Au than \dAu\ collisions); additional gain in the overall $v_2$ magnitude from hadronic rescatterings is small. The $v_2$ mass splitting is modest between primordial hadrons (arising from kinematics in the coalescence procedure); this effect is reduced if decay products are included. Despite the little gain in the overall $v_2$, a significant mass splitting is built up during the hadronic rescattering stage. 

Note that previous hadron cascade studies~\cite{Burau:2004ev,Petersen:2006vm,Zhou:2015iba}, including a recent one with free-streaming evolution coupled to a hadron cascade~\cite{Romatschke:2015dha}, have shown that the $v_2$ mass splitting can be generated by hadronic rescatterings. However, typically the overall $v_2$ magnitudes from hadronic scatterings significantly underestimate the measured $v_2$~\cite{Petersen:2006vm,Zhou:2015iba,Romatschke:2015dha}, while the study that roughly reproduces the $v_2$ magnitudes at low $\pt$ has used hadron degrees of freedom at very high energy densities~\cite{Burau:2004ev}. On the other hand, the overall $v_2$ in this multi-phase study is mostly generated by partonic rescatterings at high energy densities~\cite{He:2015hfa}, while the $v_2$ mass splitting mostly comes from the later hadronic scatterings. In addition, our model has already be shown to reasonably reproduce particle yields, $\pt$ spectra, and $v_2$ of low-$\pt$ pions and kaons in Au+Au collisions~\cite{Lin:2014tya}.

{\em Summary.}
Previous studies have shown that the measured azimuthal anisotropies $v_n$ in heavy ion as well as small system collisions at low $\pt$ can be well described by both hydrodynamics and a multi-phase transport (AMPT) model. The mass splitting of $v_n$ is considered as strong evidence for hydrodynamic collective flow. However, a recent study indicates that the major source of $v_n$ in AMPT is an anisotropic escape mechanism. In particular, it is the {\em only} source of $v_n$ in the \phirndm\ test of AMPT.

Here we have studied the development of the $v_2$ mass splitting at different stages of nuclear collisions in AMPT. We find that, while the $v_2$ amplitude is dominantly developed during the parton cascade stage, the  $v_2$ mass splitting is relatively small right after hadronization, especially when resonance decays are included. This mass splitting before hadronic rescatterings is produced by dynamics in the coalescence process such as kinematics. We demonstrate that the majority of the mass splitting is developed in the hadronic rescattering stage, although the gain in the overall charged hadron $v_2$ magnitude is small. These qualitative conclusions are the same as those from hybrid models that couple hydrodynamics to a hadron cascade. In addition, we found no qualitative difference between Au+Au collisions and \dAu\ collisions. 
We conclude that the mass splitting of $v_2$ cannot be considered as a unique signature of hydrodynamic collective flow, and the $v_2$ mass splitting cannot distinguish whether the elliptic flow is generated mainly from hydrodynamics or the anisotropic parton escape.

{\em Acknowledgments.}
We thank Dr.~Jurgen Schukraft for discussions.
This work is supported in part by US~Department of Energy Grant No.~DE-FG02-88ER40412 (LH,FW,WX) and No.~DE-FG02-13ER16413 (DM). HL acknowledges financial support from the China Scholarship Council.


\end{document}